\documentclass[pre,twocolumn]{revtex4-1}
\pdfoutput=1

\usepackage{graphicx}

\begin{document}

\title{Analyzing complex networks through correlations in centrality
  measurements}

\author{Jos\'e Ricardo Furlan Ronqui}\email{jose.ronqui@usp.br}
\author{Gonzalo Travieso}\email{gonzalo@ifsc.usp.br}
\affiliation{Instituto de F\'{\i}sica de S\~ao Carlos, Universidade de
  S\~ao Paulo, Brazil}

\begin{abstract}
  Many real world systems can be expressed as complex networks of
  interconnected nodes.  It is frequently important to be able to
  quantify the relative importance of the various nodes in the
  network, a task accomplished by defining some centrality measures,
  with different centrality definitions stressing different aspects of
  the network.  It is interesting to know to what extent these
  different centrality definitions are related for different networks.
  In this work, we study the correlation between pairs of a set of
  centrality measures for different real world networks and two
  network models.  We show that the centralities are in general
  correlated, but with stronger correlations for network models than
  for real networks.  We also show that the strength of the
  correlation of each pair of centralities varies from network to
  network.  Taking this fact into account, we propose the use of a
  \emph{centrality correlation profile}, consisting of the values of
  the correlation coefficients between all pairs of centralities of
  interest, as a way to characterize networks.  Using the yeast
  protein interaction network as an example we show also that the
  centrality correlation profile can be used to assess the adequacy of
  a network model as a representation of a given real network.
\end{abstract}

\maketitle

\section{Introduction}
\label{sec:intro}

Important aspects of many systems can be represented by complex
networks~\cite{Barabasi02,Dorogovtsev02,Newman03,Newman10,Costa11}.
As the nodes frequently differ with respect to their essentiality,
since the beginning of the study of networks the quantification of the
importance of their nodes has been receiving attention, as can be
seen, e.g.\ in
Refs.~\cite{Freeman78,Bolland88,Friedkin91,Faust97,Jeong01,Borgatti06,Manimaran09,Wang11}.
There are different ways to quantify the importance, or centrality, of
a node, and therefore a large number of measures used for this
purpose, with new centrality measures being constantly proposed for
use in new applications or to achieve better results in old ones, see
e.g.,
Refs.~\cite{Bonacich87,Stephenson89,Rothenberg95,Valente98,Newman05,Estrada05,Borgatti05,Costa07,Estrada08,Guimera05,Koschutzki05,Koschutzki08,Kitsak10,Pauls12,Avrachenkov13,Molinero13,Campiteli13,Benzi13}.

Although based on different definitions, the various node centralities
are, in real networks, correlated: important nodes using one of the
definitions are frequently also important using others.  For example,
nodes with high degree have also high closeness
centrality~\cite{Bolland88}.  Some papers already analysed those
correlations, while others do correlation analysis when proposing new
centralities~\cite{Bolland88,Valente08,Ochab12,Pauls12,Avrachenkov13,Iyer13,Campiteli13,Benzi13}.
Nonetheless, there are nodes with high value for one centrality and
low value for another, and the correlations are not the same for all
networks, as will be shown below.

In this work we systematically study the correlations between all
pairs of a set of centralities using some real-world networks and two
network models.  We find that the correlations are generally strong,
but there are marked differences between correlations in real network
and models, and also among different real networks.  We therefore
suggest the use of such correlations as a way to characterize
networks.

This paper is organized as follows. In Section~\ref{sec:conc} we
present the considered network centralities and the real networks and
models used. In Section~\ref{sec:res} we present scatter plots of the
centrality values for some pairs of centralities
(Section~\ref{sec:corr}) and show that the relations are mostly power
law; taking into account this power-law behavior, we present Pearson
correlations for the \emph{logarithms} of the centralities, which will
give us a measure of how closely the centralities are related by a
power law.  Next (Section~\ref{sec:prof}) we present the concept of
\emph{centrality correlation profile} to characterize networks.  This
is followed by showing (Section~\ref{sec:rewire}) that this centrality
correlation profile can be used to distinguish the real networks from
their randomly rewired counterparts, as well as from the network
models.  Finally (Section~\ref{sec:models}) we propose the use of
this profile to evaluate models for a given real network, using as an
example a model for protein-protein interaction networks.

\section{Basic concepts and datasets}
\label{sec:conc}

We are considering only undirected, unweighted networks without
multiple edges or self connections.  In this case, the network can be
represented by a symmetric \emph{adjacency matrix} $\mathbf{A}$ whose
$N \times N$ elements (where $N$ is the number of nodes) $A_{ij}$ are
$1$ if nodes $i$ and $j$ are connected and $0$ otherwise.  Some
networks are in fact weighted, as described below, but we disregard
the weights.  We also drop self and multiple connections when
present.  When a network has more than one connected component, we
consider only the nodes in the largest component.

An important concept is that of \emph{shortest paths}.  A path is a
sequence of nodes where each two subsequent nodes are directly
connected and no node is repeated in the path.  A shortest path
between nodes $i$ and $j$ is a path starting at node $i$, ending at
node $j$ and with the smallest possible number of intermediate nodes
in the path.

\subsection{Node centralities}
\label{sec:centr}

There is a large number of centrality measures in the literature.  For
brevity, we will work with some of them, including the most used ones,
instead of trying to be comprehensive.  Although other centralities
can be important in many applications, the methodology employed here
could be, if needed, easily extended to include other centralities.
Furthermore, we use PCA (see Sec.~\ref{sec:prof}) to automatically
compensate for possible redundancies among the centralities, and the
results show that using other centralities would not contribute
significantly for the considered networks (as enough discrimination is
already achieved).  It is plausible that other centralities could be
necessary for a different dataset.

\paragraph{Degree centrality}

This centrality quantifies the importance of a node counting its
number of connections.  Using the adjacency matrix, the degree of node
$i$, represented as $k_i$ is computed as
\begin{equation}
  \label{eq:degree}
  k_i = \sum_{j=1}^{N}A_{ij}.
\end{equation}
Here we use a normalized degree centrality, given by dividing the
degree by the maximum possible degree:
\begin{equation}
  \label{eq:degreecentrality}
  \tilde{k}_i = \frac{k_i}{N-1}.
\end{equation}

\paragraph{Eigenvector centrality}
Just counting the number of connections, as done in the degree
centrality, can give a distorted view of the importance of a node,
because it does not quantify the importance of its neighbors.  In
principle, the importance of the neighbors should be considered when
accessing the importance of a node.  If $v_i$ is the importance of node
$i$, we can compute it in a self-consistent way through
\begin{equation}
  \label{eq:eigen}
  v_i = \frac{1}{\lambda} \sum_j A_{ij}v_j.
\end{equation}
where $\lambda$ must be chosen appropriately.  In vector form we have:
\begin{equation}
  \label{eq:eigenvector}
  \lambda \mathbf{v} = \mathbf{A}\mathbf{v},
\end{equation}
which tells us that $\mathbf{v}$ is an eigenvector of the adjacency
matrix and $\lambda$ the corresponding eigenvalue.  In fact, we use
the eigenvector associated with the \emph{largest} eigenvalue of the
adjacency matrix, and the \emph{eigenvector centrality} of node $i$
is the $i$-th entry in this eigenvector.

\paragraph{Closeness centrality}
It is also possible to take the word ``centrality'' more literally and
search for nodes that are central in the sense of being in average
closer to the other nodes.  If $d_{ij}$ is the shortest path
distance between nodes $i$ and $j$ we can compute the closeness
centrality of node $i$ as
\begin{equation}
  \label{eq:closeness}
  c_i = \frac{1}{\sum_j d_{ij}}.
\end{equation}

\paragraph{Betweenness centrality}
Assuming pairs of nodes in the network must interact, if they are not
directly connected the interaction must go through intermediary nodes.
A node is important in the betweenness centrality sense if it must be
used as an intermediary for many pairs of nodes (under the assumption
that the interactions always follow a shortest path, that is, a path
with minimum number of intermediaries).

The betweenness centrality of node $i$, represented as $b_i$, is
computed by the expression:
\begin{equation}
  \label{eq:between}
  b_i = \sum_{j,k} \frac{n(j,i,k)}{n(j,k)},
\end{equation}
where $j \neq i \neq k$, $n(j,k)$ is the number of shortest paths from
$j$ to $k$ and $n(j,i,k)$ is the number of shortest paths from $j$ to
$k$ that pass through $i$.

\paragraph{Current flow betweenness centrality}

Betweenness centrality takes into account only the shortest paths from
a node $j$ to another node $k$.  It is possible for the nodes to
interact through other paths.  This is taken into account by the
centrality measure based on computing the current flow through the
network elements supposing that each link is a resistor (with a value
of 1 for unweighted networks) considering all possible sources and
drains for the current.  This is equivalent to counting the number of
times a random walk from a node $j$ to node $k$ passes through a given
node $i$, for all pair $(j,k)$ (but canceling back-and-forth movements
of the walker that do not contribute to a net movement toward the
target)~\cite{Newman05,Brandes05} and is therefore also called random
walk betweenness centrality.

\paragraph{Current flow closeness}

Also know as information centrality~\cite{Brandes05}, this measure
first proposed in Ref.~\cite{Stephenson89}, is a generalization of the
closeness centrality in the same lines than the current flow (or
random walk) betweeness centrality is a generalization of the
shortest-path betweenness centrality: by considering alternate paths
from a node to other nodes instead of just the shortest path.

\paragraph{Subgraph centrality}

This measure takes into account the participation of a node in
subgraphs, given larger weight for smaller subgraphs~\cite{Estrada05}.
Closed walks starting and ending in a node $i$ are counted and
weighted with the inverse factorial of their size.  With the chosen
weighting, the values can be efficiently computed using the spectral
decomposition of the adjacency matrix.  If $\lambda_j$ are the
eigenvalues and $\mathbf{v}_j$ are the respective eigenvectors, the
subgraph centrality $s_i$ of node $i$ can be computed using the
expression
\begin{equation}
  \label{eq:subgraph}
  s_i = \sum_{j=1}^{N}{(v_j^i)}^2e^{\lambda_j},
\end{equation}
where $v_j^i$ is the $i$-th element of eigenvector $\mathbf{v}_j$.

\subsection{Networks and network models}
\label{sec:net}

We work here with the following available network datasets: Zachary's
karate club (represents friendship between 34 members of a karate
club)~\cite{Zachary77}; dolphin social network (frequent association
between 62 dolphins)~\cite{lusseau03}; high-energy theory
collaboration (coauthorship in preprints on the hep-th section in
arXiv.org)~\cite{Newman01:pnas,Newman01:pre:1,Newman01:pre:2}; network
science collaborations (coauthorship in network science
papers)~\cite{Newman06}; books about US politics published around 2004
and sold on Amazon.com (edges show frequent co-purchase)~\cite{Krebs};
power grid (topology of the power grid of the Western States of the
USA)~\cite{Watts98}.  Table~\ref{tab:meas} shows some measurements for
the networks.

 \begin{table*}[!htpb]
  \centering
  \caption{Topological measurements for the (largest components of
    the) networks: karate club (KT), dolphins (DP), high-energy
    physics collaboration (HP), network science collaboration (NS),
    political books (PB), and power grid (PW).  $N$: number of nodes,
    $m$: number of edges, $\langle k \rangle$: average degree,
    $\langle k^{2} \rangle$: variance of the degree distribution,
    $\triangle_{\mathrm{local}}$: average local clustering coefficient,
    $\triangle_{\mathrm{transitivity}}$: global clustering coefficient
    (transitivity), $\ell$: average shortest path length, $E$:
    efficiency, $r$: assortativity.}
  \begin{ruledtabular}
  \begin{tabular}{lcccccc}
    Measures                &     KT   &     DP   &     HP   &     NS   &     PB   &     PW    \\ 
    \hline
    $N$                           &     34   &     62   &     5835 &     379  &     105  &     4941  \\ 
    $m$                           &     78   &     159  &    13815 &     914  &     441  &     6594  \\ 
    $\langle k \rangle$         &     4.59 &     5.13 &     4.74 &     4.82 &     8.4  &      2.67 \\ 
    $\langle k^{2} \rangle$     &    35.65 &    34.90 &    43.19 &    38.69 &   100.25 &     10.33 \\ 
    $\triangle_{\mathrm{local}}$         &     0.57 &     0.26 &     0.51 &     0.74 &     0.49 &      0.08 \\ 
    $\triangle_{\mathrm{transitivity}}$  &     0.26 &     0.31 &     0.28 &     0.43 &     0.35 &      0.10 \\ 
    $\ell$          &     2.34 &     3.30 &     7.03 &     6.03 &     3.05 &     18.99 \\ 
    $E$                  &     0.49 &     0.38 &     0.16 &     0.20 &     0.40 &      0.06 \\ 
    $r$               &    -0.48 &    -0.04 &     0.19 &    -0.08 &    -0.13 &      0.00 \\ 
  \end{tabular}
  \end{ruledtabular}
\label{tab:meas}
\end{table*}

Our emphasis is showing results for real networks.  We therefore
include only two simple models for comparison, the Erd\H{o}s-R\'enyi
(ER) random graphs~\cite{Erdos59} and the Barab\'asi-Albert (BA)
scale-free networks~\cite{Barabasi97}.  The method used here could be
applied for other models (as done in Section~\ref{sec:models}), if
appropriate.

\section{Results and discussion}
\label{sec:res}

Given a network, we compute the centralities of each of its nodes and
search for correlations between pairs of centralities, with each node
in the network corresponding to a data point.

\subsection{Correlations}
\label{sec:corr}

\begin{figure*}[!hptb]
  \centering
  \includegraphics[width=0.32\textwidth]{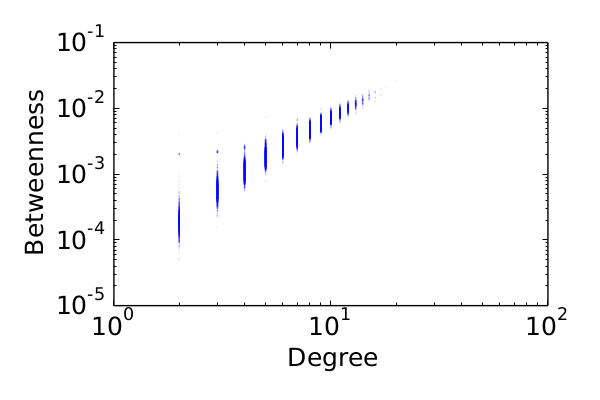}
  \includegraphics[width=0.32\textwidth]{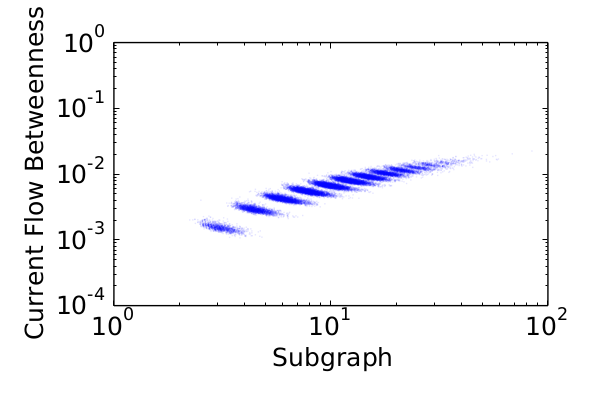}\\
  \includegraphics[width=0.32\textwidth]{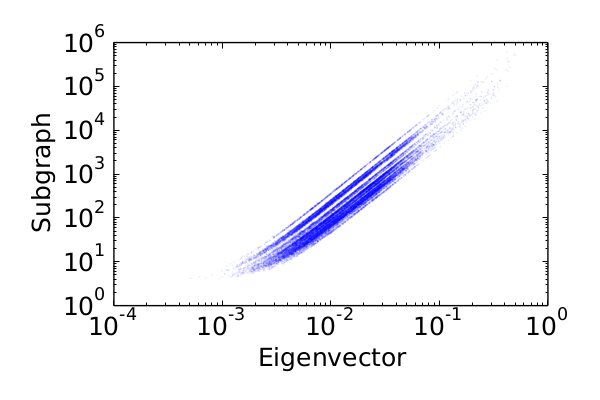}
  \includegraphics[width=0.32\textwidth]{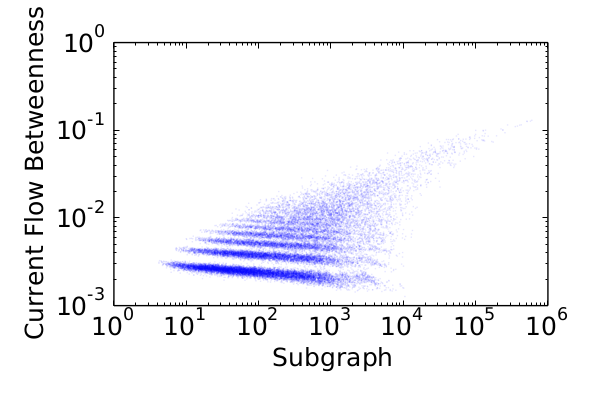}
  \caption{Scatter plots for some pairs of centralities for the
    Erd\H{o}s-R\'enyi (top) and Barab\'asi-Albert (bottom) network
    models. Left: Best correlations (excluding closeness related
    cases, see text); Right Worst
    correlations.}\label{fig:scattermodels}
\end{figure*}

\begin{figure*}[!hptb]
  \centering
  \includegraphics[width=0.32\textwidth]{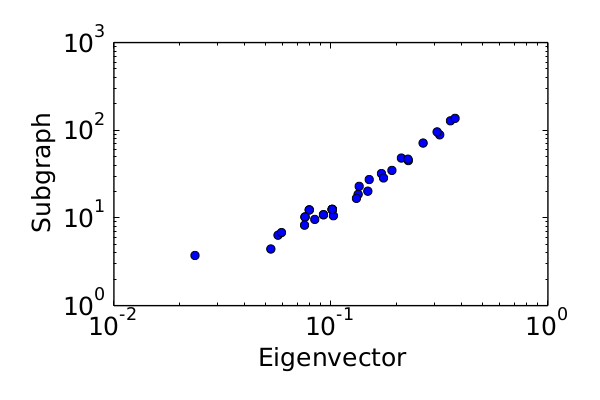}
  \includegraphics[width=0.32\textwidth]{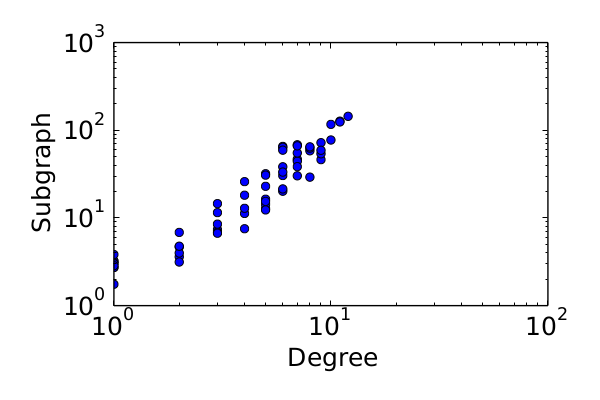}
  \includegraphics[width=0.32\textwidth]{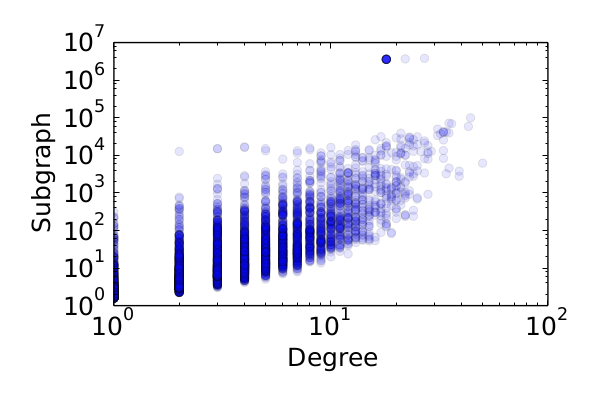}\\
  \includegraphics[width=0.32\textwidth]{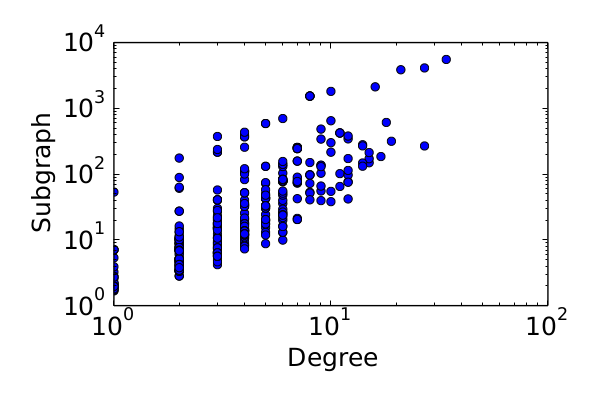}
  \includegraphics[width=0.32\textwidth]{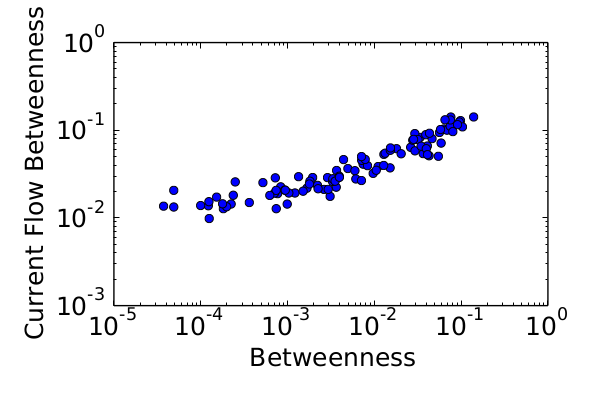}
  \includegraphics[width=0.32\textwidth]{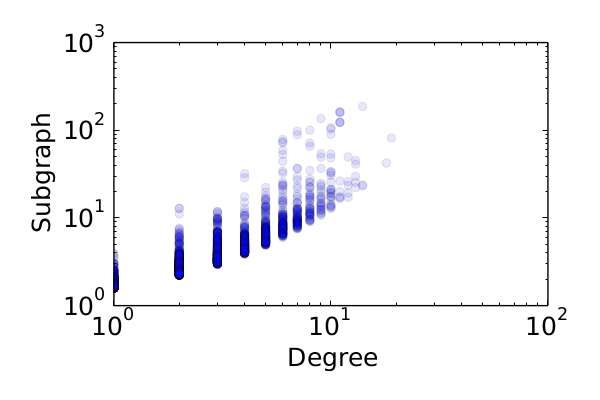}
  \caption{Scatter plots for the best correlations between pairs of
    centralities for the real networks (excluding closenness related
    cases).  Top, left to right: karate, dolphins, and high-energy
    physics; bottom, left to right: network science, political books,
    and power grid.}\label{fig:scatterreal-good}
\end{figure*}

\begin{figure*}[!hptb]
  \centering
  \includegraphics[width=0.32\textwidth]{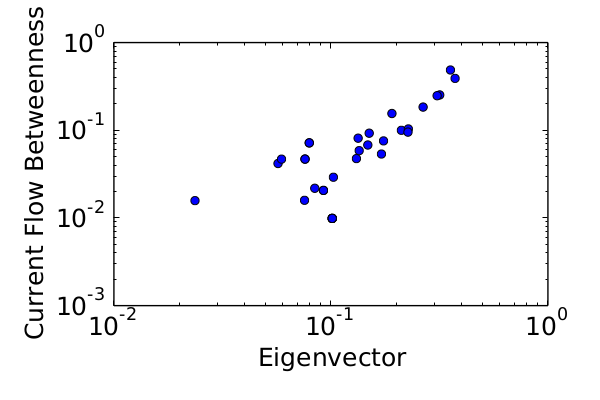}
  \includegraphics[width=0.32\textwidth]{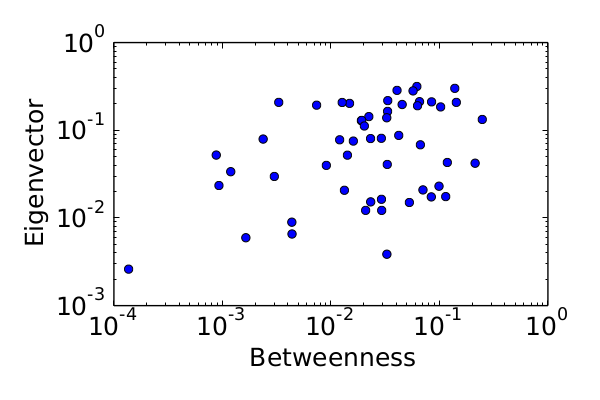}
  \includegraphics[width=0.32\textwidth]{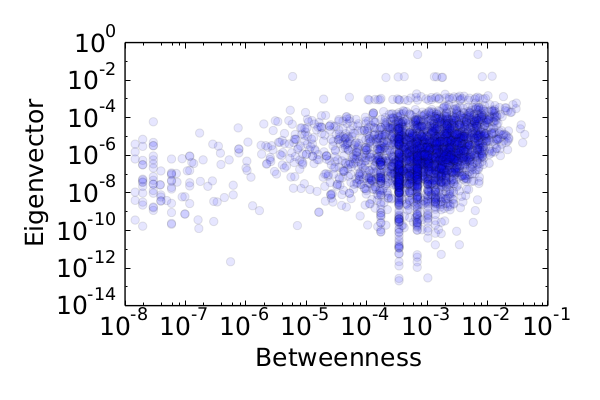}\\
  \includegraphics[width=0.32\textwidth]{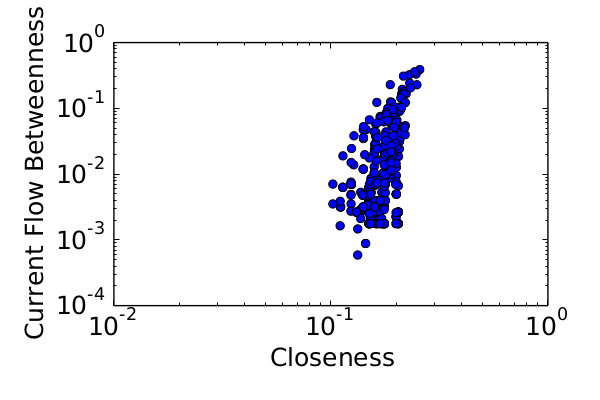}
  \includegraphics[width=0.32\textwidth]{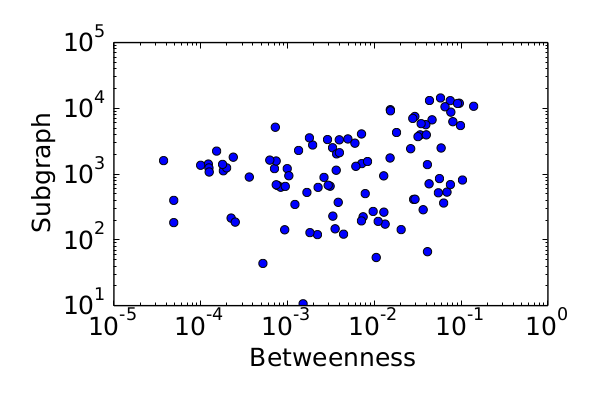}
  \includegraphics[width=0.32\textwidth]{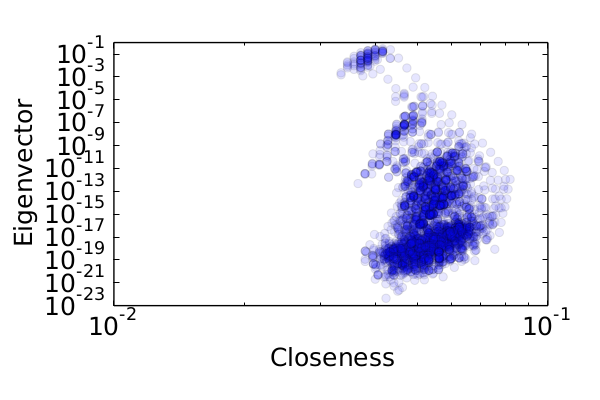}
  \caption{Scatter plots for the worst correlations between pairs of
    centralities for the real networks.  Top, left to right: karate,
    dolphins, and high-energy physics; bottom, left to right: network
    science, political books, and power grid.
  }\label{fig:scatterreal-bad}
\end{figure*}

Figure~\ref{fig:scattermodels} shows scatterplots for some of the
pairs of centralities for the network models, while the plots for the
real networks are presented in Figure~\ref{fig:scatterreal-good} (best
correlations) and Figure~\ref{fig:scatterreal-bad} (worst
correlations).  We do not show closeness or current flow closeness
centralities results for best cases as these measurements span a
limited range, which limits the significance of a good correlation in
a log log plot.  Due to space limitations, we show only two of the
pairs with largest and smallest correlation values, respectively, for
each network.  These plots suggest that the centralities are
correlated, with visible correlations even in the weakest cases for
some networks, and that the correlations are close to a power law,
specially for high values of centralities.

To quantify how closely two measurements are related by a power law,
we use log-log plots and compute the Pearson correlation coefficient
between the logarithms of the values of the centralities.  Similar
results were also found when using the Pearson and Spearman
correlations of the measurements (without the logarithms).  The use of
the logarithms here is to emphasize possible power-laws, as observed
in Figures~\ref{fig:scattermodels} and~\ref{fig:scatterreal-bad}.

Table~\ref{tab:corr} shows the values of the Pearson coefficients for
all pairs of logarithms of the centralities in the networks studied.
For the ER and BA models, 50~networks (for each model) of
1000~vertices and average degree~6 where used.  We can see that the
centralities have, in general, large values of the Pearson
coefficient, which in our case implies proximity to a power law
relation.  With the exception of a small negative coefficient between
closeness and eigenvector centralities for the power grid network, all
coefficients are positive.  This means that nodes that are important
with respect to one definition are, in general, also important
according to other definitions.  It can be seen that the network
models studied present larger coefficients than the real networks
(with the exception of the karate club network, which also has large
coefficients.\footnote{The karate club is a small network, with a few
  dominating nodes and many peripheral nodes.  This could explain the
  large values.})  Especially noticeable is the coefficient of
closeness with eigenvector centrality, which is almost perfect for the
network models (0.99 for ER and 0.98 for BA), but non-existent for the
power grid network.  It is, therefore, important to be careful when
generalizing conclusions from results using such simplified models to
real networks.  The coefficients of degree with random walk closeness
and subgraph centrality are large for all networks, with the exception
of the power grid network.  To a lesser extend, the same is true for
other pairs involving degree, betweenness and current flow
betweenness.  Other pairs have large coefficients in some networks,
but small coefficients in others.  For instance, betweenness and
subgraph centralities have large coefficients for the network models,
the dolphins and karate networks, but small values for the other
networks.  Most interesting is the case of betweenness and eigenvector
centralities: they have small coefficients for all real networks (with
the exception of the karate club network, where it is large, but
smaller than for other pairs), but large coefficients for the network
models.  This suggests that they complement each other when analising
real world networks, and reinforces our previous observation of
inadequacy of generalizing conclusions based on simple models.

\begin{table*}[!htpb]
  \centering
  \caption{Pearson coefficients for the networks: karate club (KT),
    dolphins (DP), high-energy physics collaboration (HP), network
    science collaboration (NS), political books (PB), power grid (PW),
    Erd\H{o}s-R\'{e}nyi model (ER), and Barab\'asi-Albert model (BA).
    The largest (absolute) value of correlation for each network and
    all values within a $0.05$ inclusive range are marked in bold;
    idem for the smallest (absolute) values, marked in italics.  (CF
    is an abbreviation for ``current flow''.)}
  \begin{ruledtabular}
  \begin{tabular}{lcccccccc}
    Measurements                &     KT   &     DP   &     HP   &     NS   &     PB   &     PW    &     ER   &     BA   \\ 
    \hline
    Degree/Closeness            &     0.80 &     0.75 &     0.54 &     0.26 &     0.61 &      0.22 &     0.92 &     0.67 \\ 
    Degree/Betweenness          &     0.84 &     0.73 &     0.52 &     0.58 &     0.65 &      0.47 & \bf 0.98 &     0.92 \\ 
    Degree/Eigenvector          &     0.82 &     0.63 &     0.49 &     0.21 &     0.69 &      0.16 &     0.91 &     0.62 \\ 
    Degree/Subgraph             &     0.91 & \bf 0.94 &     0.79 & \bf 0.79 &     0.83 & \bf  0.89 & \bf 0.97 &     0.62 \\ 
    Degree/CF betweenness       & \bf 0.95 &     0.70 &     0.59 &     0.51 &     0.79 &      0.46 &     0.51 & \bf 0.94 \\ 
    Degree/CF closeness         & \bf 0.95 & \bf 0.96 &     0.77 &     0.59 & \bf 0.95 &      0.35 & \bf 0.98 & \bf 0.97 \\ 
    Closeness/Betweenness       & \it 0.78 &     0.71 &     0.35 &     0.40 &     0.75 &      0.40 &     0.90 &     0.77 \\ 
    Closeness/Eigenvector       & \bf 0.92 &     0.83 & \bf 0.91 &     0.58 &     0.45 & \it -0.04 & \bf 0.99 & \bf 0.98 \\ 
    Closeness/Subgraph          &     0.89 &     0.72 &     0.71 &     0.34 &     0.47 &      0.16 & \bf 0.96 & \bf 0.97 \\ 
    Closeness/CF betweenness    &     0.82 &     0.55 &     0.48 & \it 0.14 &     0.84 &      0.50 & \it 0.43 & \it 0.54 \\ 
    Closeness/CF closeness      &     0.89 &     0.87 &     0.85 & \bf 0.76 &     0.69 &      0.75 &     0.93 &     0.70 \\ 
    Betweenness/Eigenvector     & \it 0.78 & \it 0.39 & \it 0.24 &     0.34 & \it 0.35 & \it  0.05 &     0.89 &     0.72 \\ 
    Betweenness/Subgraph        &     0.82 &     0.60 &     0.30 &     0.33 & \it 0.31 &      0.35 & \bf 0.95 &     0.70 \\ 
    Betweenness/CF betweenness  &     0.90 & \bf 0.93 &     0.69 &     0.71 & \bf 0.93 &      0.55 & \bf 0.98 &     0.91 \\ 
    Betweenness/CF closeness    &     0.85 &     0.71 & \it 0.29 &     0.39 &     0.62 &      0.34 & \bf 0.96 & \bf 0.93 \\ 
    Eigenvector/Subgraph        & \bf 0.97 &     0.70 &     0.71 &     0.46 &     0.82 &      0.32 & \bf 0.95 & \bf 0.96 \\ 
    Eigenvector/CF betweenness  & \it 0.74 &     0.61 &     0.40 & \it 0.15 &     0.45 &      0.10 & \it 0.44 & \it 0.49 \\ 
    Eigenvector/CF closeness    & \bf 0.92 &     0.76 &     0.78 &     0.65 &     0.72 & \it  0.06 &     0.92 &     0.66 \\ 
    Subgraph/CF betweenness     &     0.82 &     0.55 &     0.41 &     0.34 &     0.46 &      0.37 & \it 0.40 & \it 0.49 \\ 
    Subgraph/CF closeness       & \bf 0.96 &     0.90 &     0.77 &     0.71 &     0.87 &      0.29 & \bf 0.94 &     0.65 \\ 
    CF betweenness/CF Closeness &     0.91 &     0.76 &     0.57 &     0.39 &     0.77 &      0.70 &     0.58 & \bf 0.93 \\ 
  \end{tabular}
  \end{ruledtabular}
\label{tab:corr}
\end{table*}

\subsection{Correlation profile of networks}
\label{sec:prof}

\begin{figure}[!hptb]
  \centering
  \includegraphics[width=0.9\columnwidth]{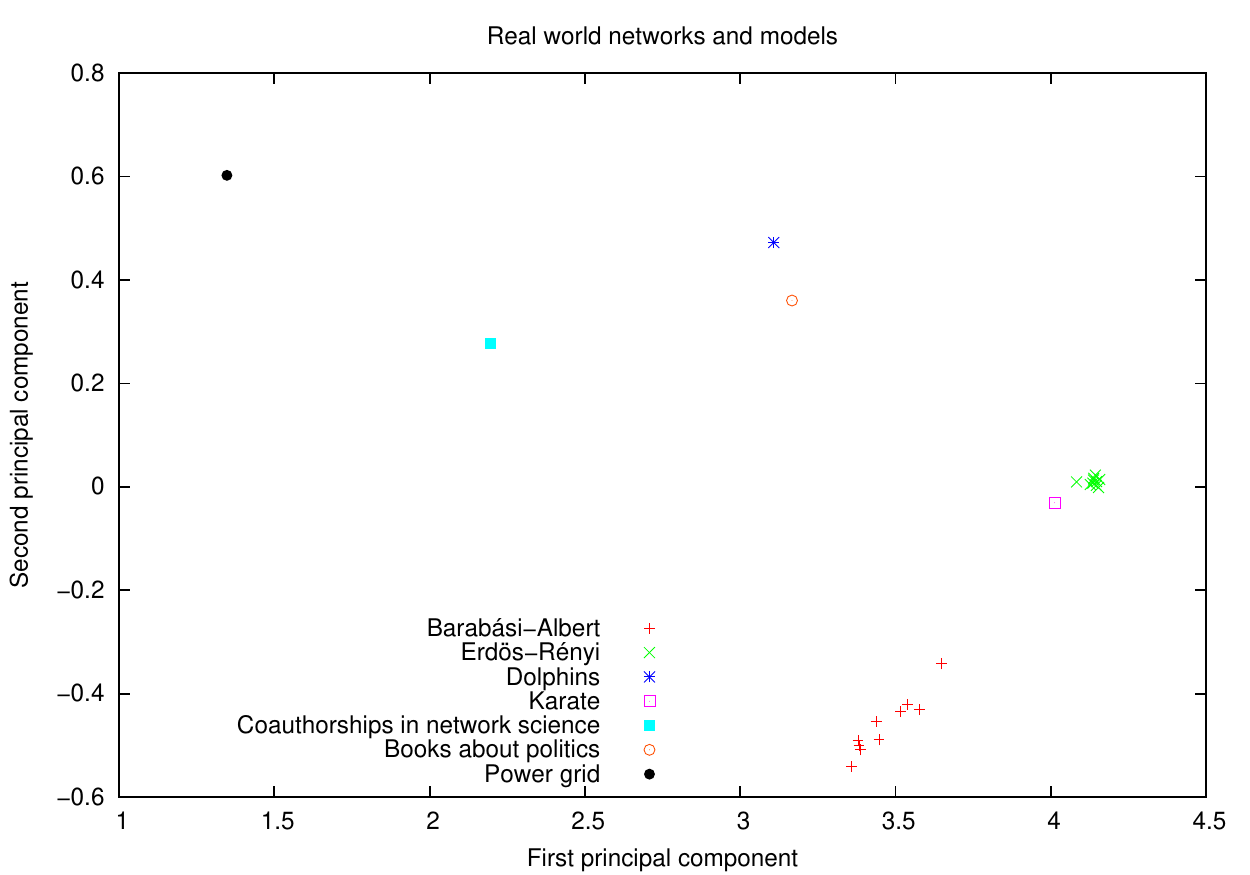}  
  \caption{PCA projection of the centrality correlation profile space
    for the networks used in this work.}\label{fig:profall}
\end{figure}

These results suggest that each network or network model has a
specific profile of correlations between centrality measurements.  We
call this the \emph{centrality correlation profile} of the network.
To show that this profile can be used to characterize the networks,
Figure~\ref{fig:profall} plots a two-dimensional projection of the
real world networks from the space defined by the centrality
correlation profile using principal component analysis
(PCA)~\cite{Abdi10}.  Each network is a point in a 21-dimensional
space defined by the values of the Pearson correlation between the
logarithms of the seven considered measurements.  The points are
projected to the two principal components for visualization.  Note how
the networks generated by the same model are clustered in small
regions, while the different real networks or models are spread
through the graph.  The only exception is the small karate club
network, which is close to the ER model cluster.

\subsection{Comparison with random rewiring}
\label{sec:rewire}

\begin{figure*}[!hptb]
  \centering
  \includegraphics[width=0.48\textwidth]{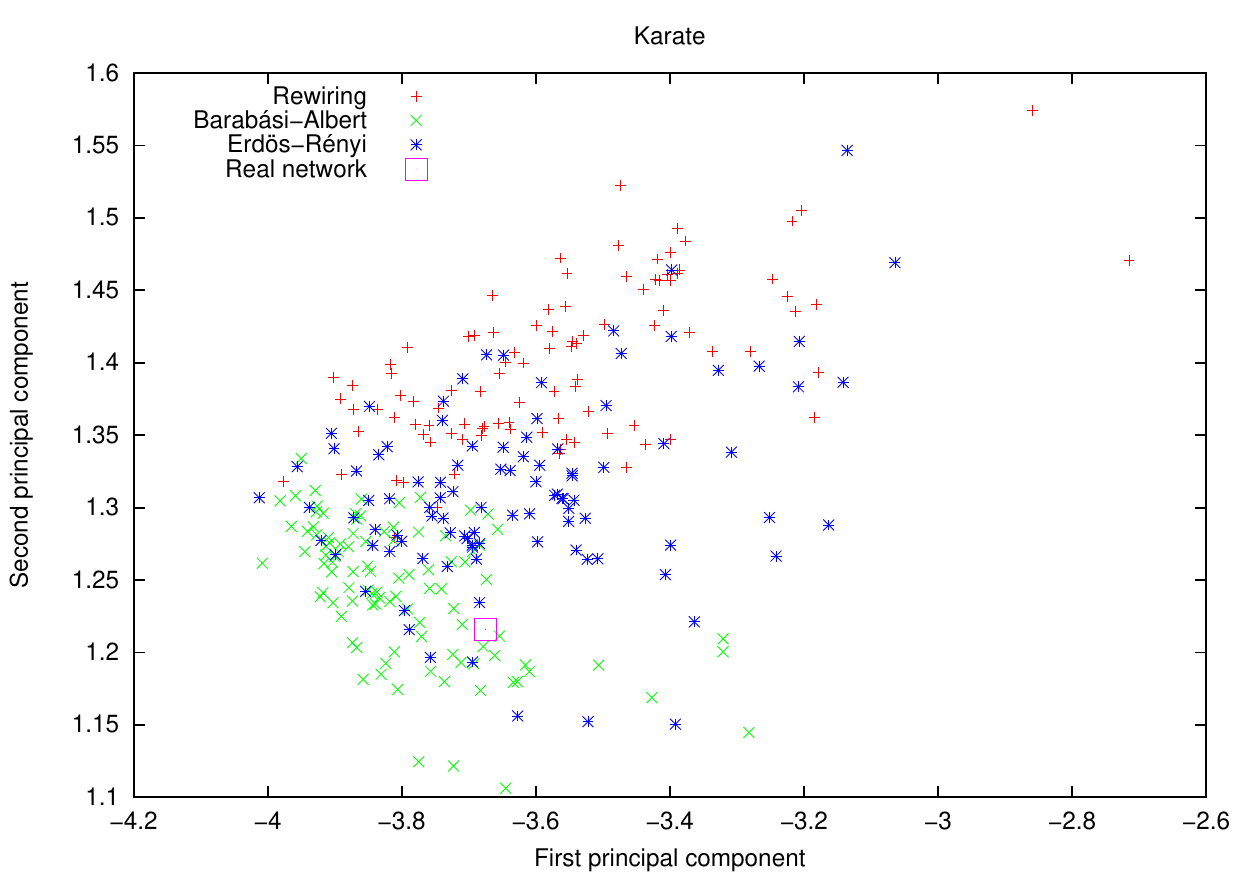}  
  \includegraphics[width=0.48\textwidth]{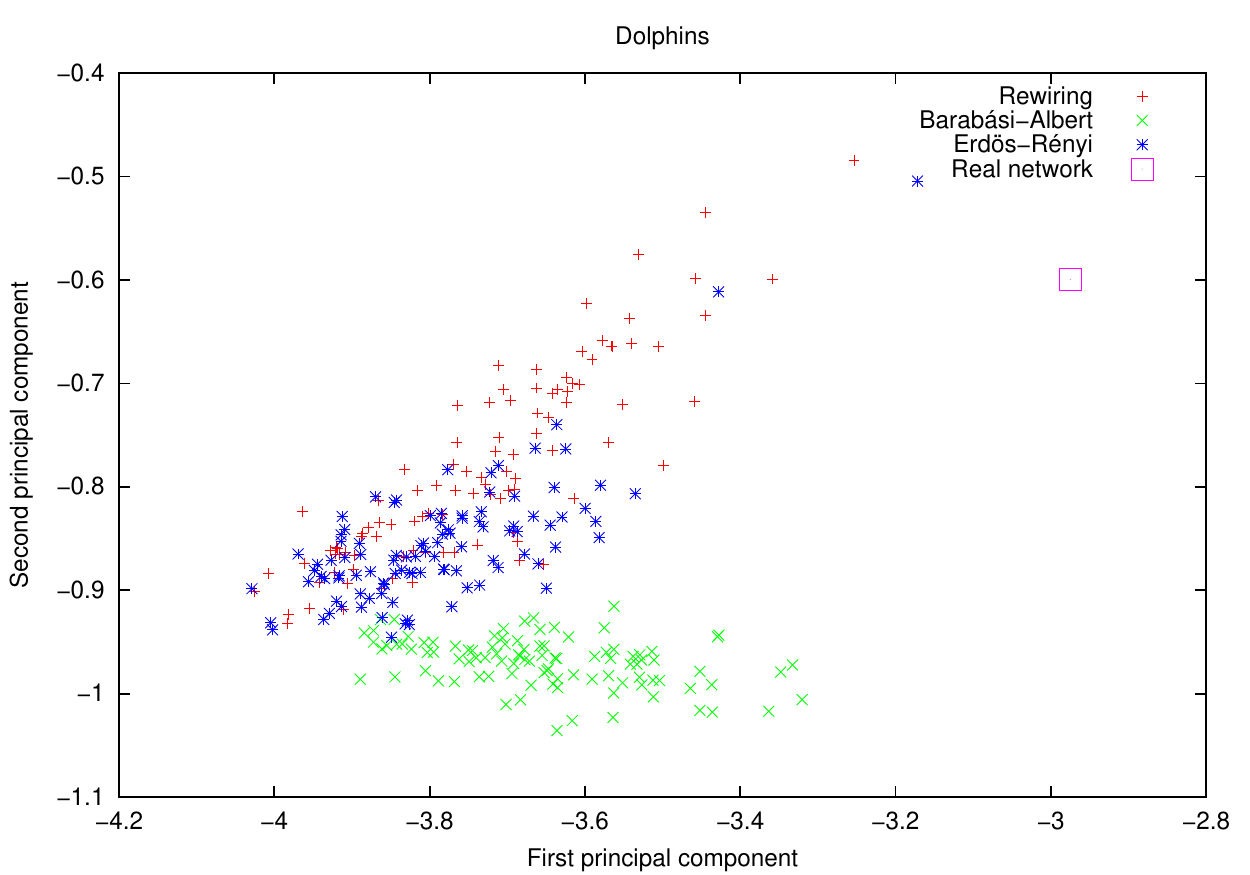}  \\
  \includegraphics[width=0.48\textwidth]{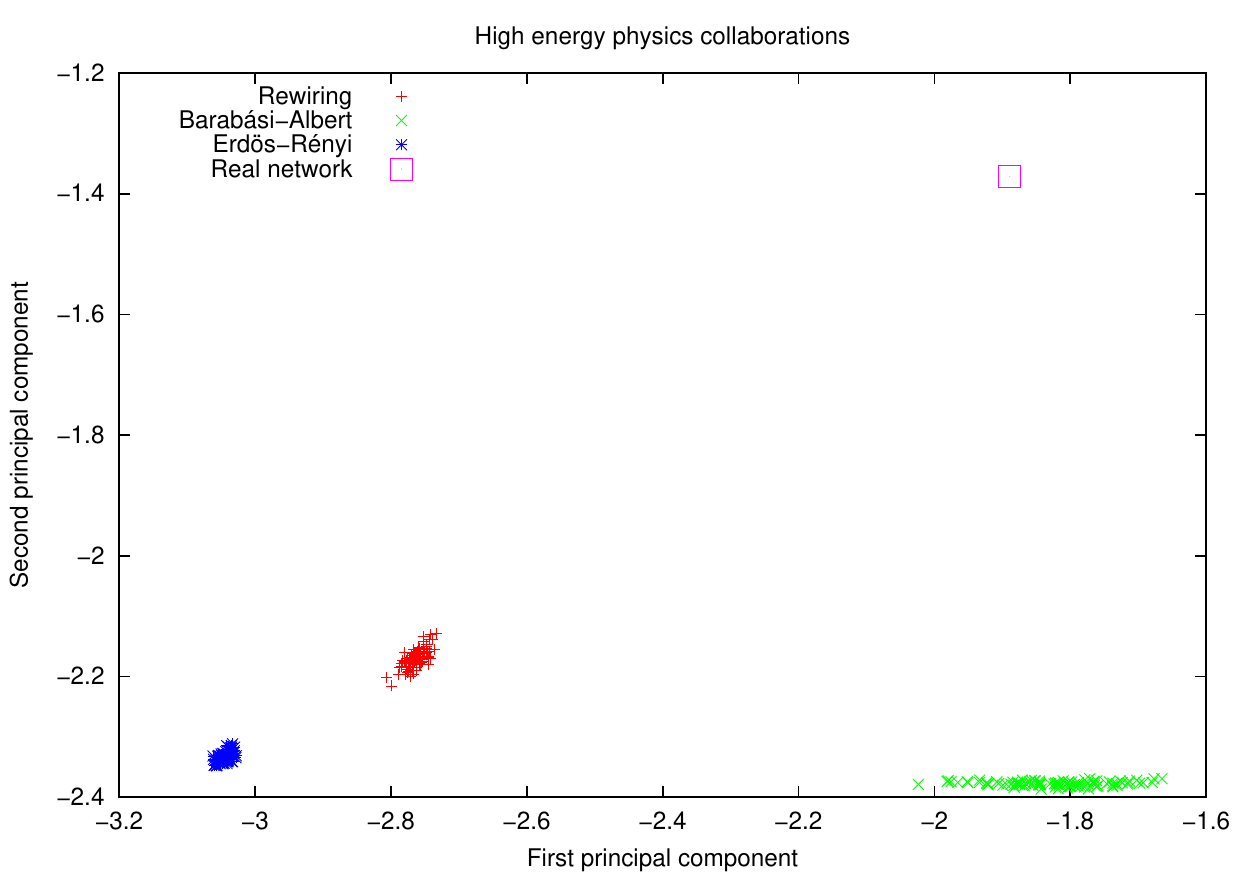}
  \includegraphics[width=0.48\textwidth]{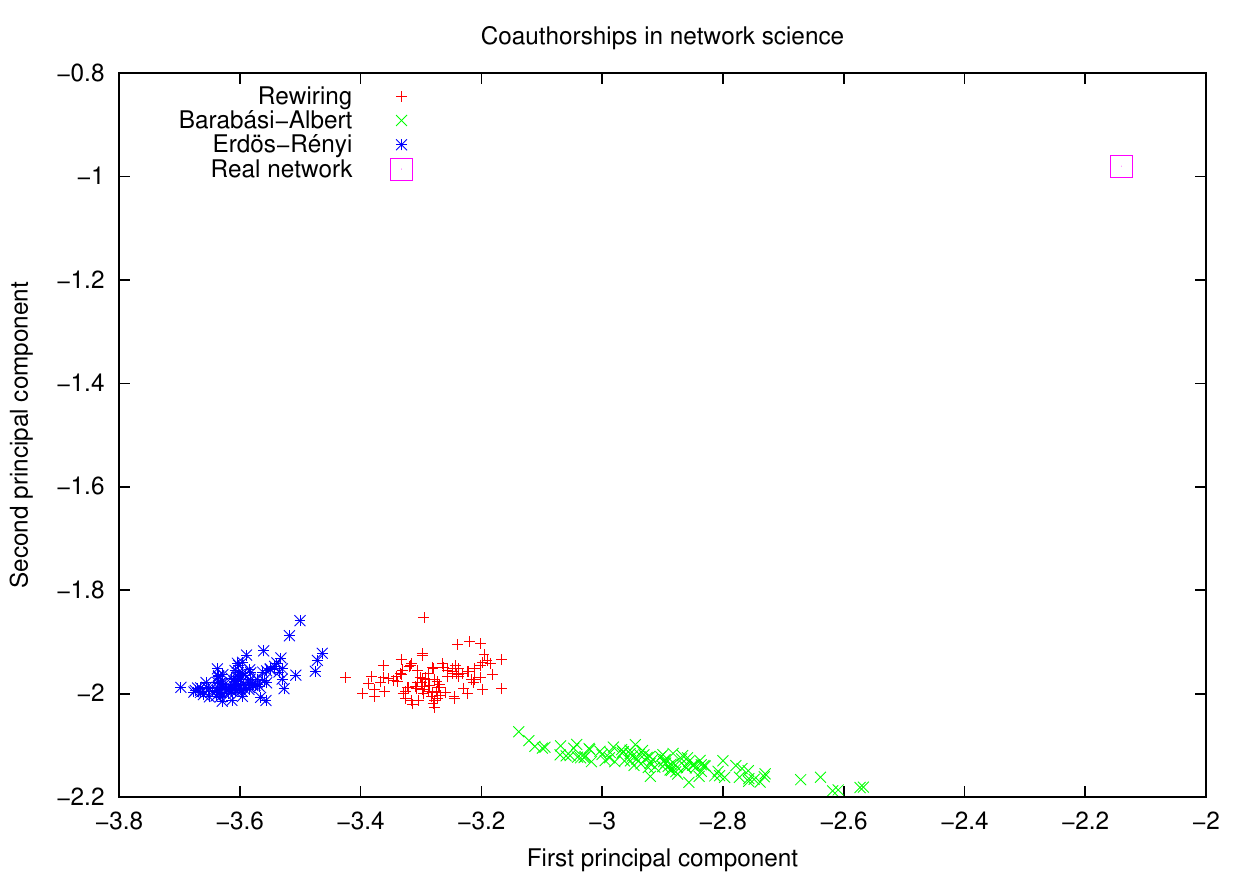}  \\
  \includegraphics[width=0.48\textwidth]{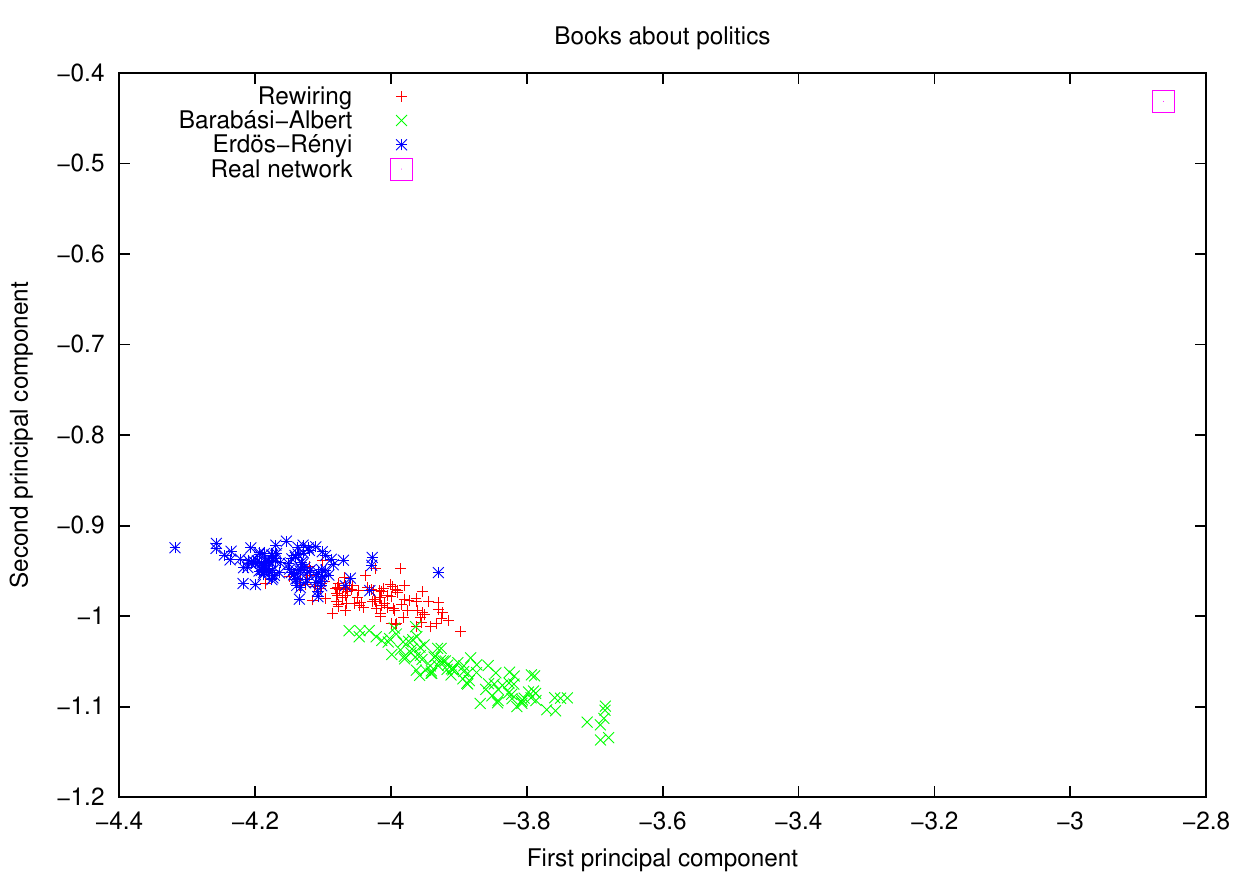}  
  \includegraphics[width=0.48\textwidth]{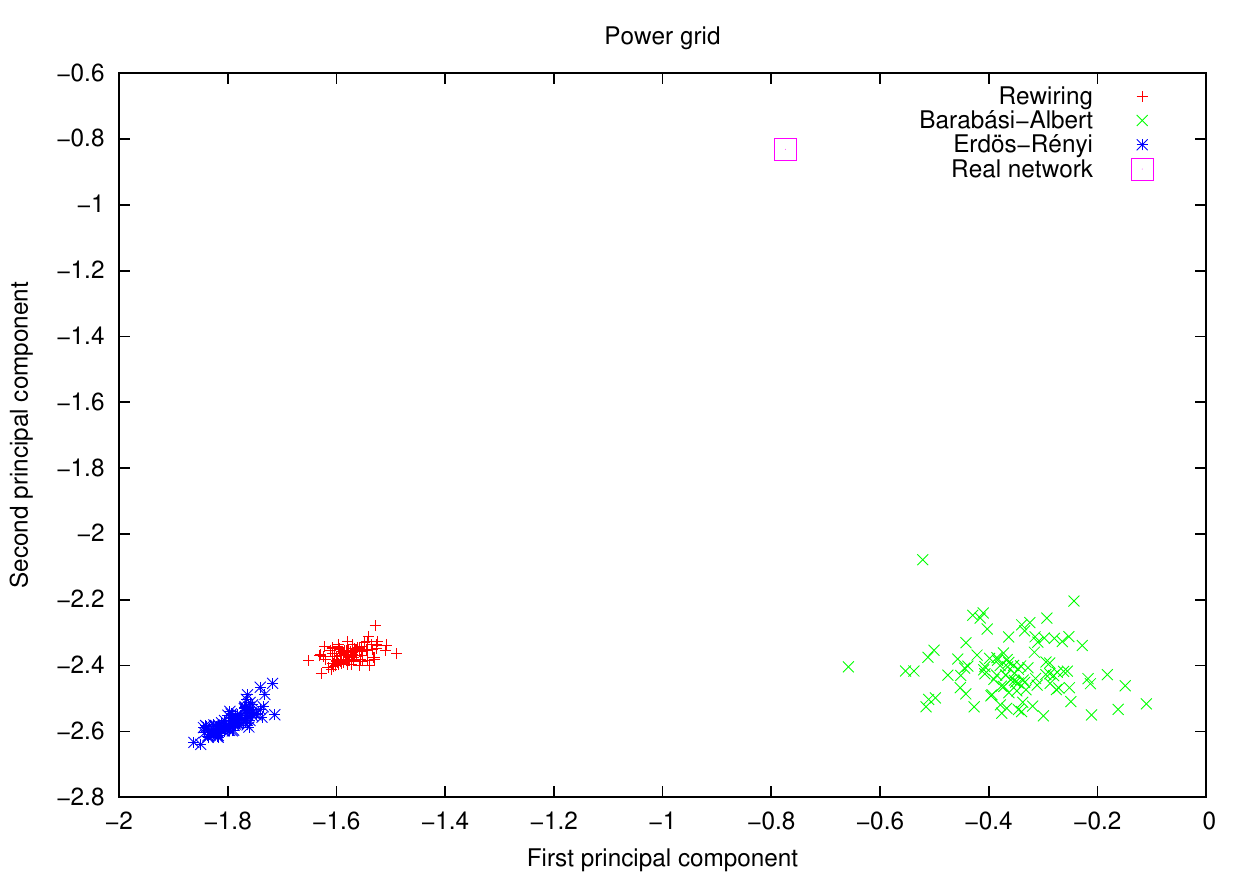}  
  
  \caption{PCA projections comparing the real networks with rewired
    random version with the same degree sequence.  From top left to
    bottom right: karate, dolphins, high-energy physics, network
    science, political books, and power grid.}\label{fig:rewire}
\end{figure*}

In our next experiment we generate, for each real network, 100 random
networks with the same degree sequence through link
rewiring~\cite{Milo02}.  In this method, a pair of edges is randomly
chosen, the original edges are removed and substituted by two new
edges among the same vertices; the process is repeated a certain
number of times.  Each rewired network is generated by a number of
random rewirings equal to 100 times the number of edges in the
original network.  We also include, for comparison, 100 networks each
for the ER and BA models with the same number of nodes and similar
average degrees.  We compute the centrality correlation profiles of
all networks and generate a two-dimensional PCA projection.  The
results are shown in Figure~\ref{fig:rewire}.  With the exception of
the karate club network, where the real network is inside the region
of the rewired networks, we can distinguish the networks from their
random rewirings and the two other models.  This stresses the fact
that, although there is generally a strong correlation between the
various centralities, there is also important information in the
specific wiring pattern of real networks, resulting in distinct
correlation profiles.  It is also interesting to note that the
randomly rewired networks are sometimes closer to the ER, sometimes to
the BA networks, but always closer to the models than the
corresponding real network, supporting the assertion that the
centrality correlation profile is characteristic of the specific
network.

\subsection{Evaluating models with the correlation profile}
\label{sec:models}

\begin{figure}[!hptb]
  \centering
  \includegraphics[width=0.96\columnwidth]{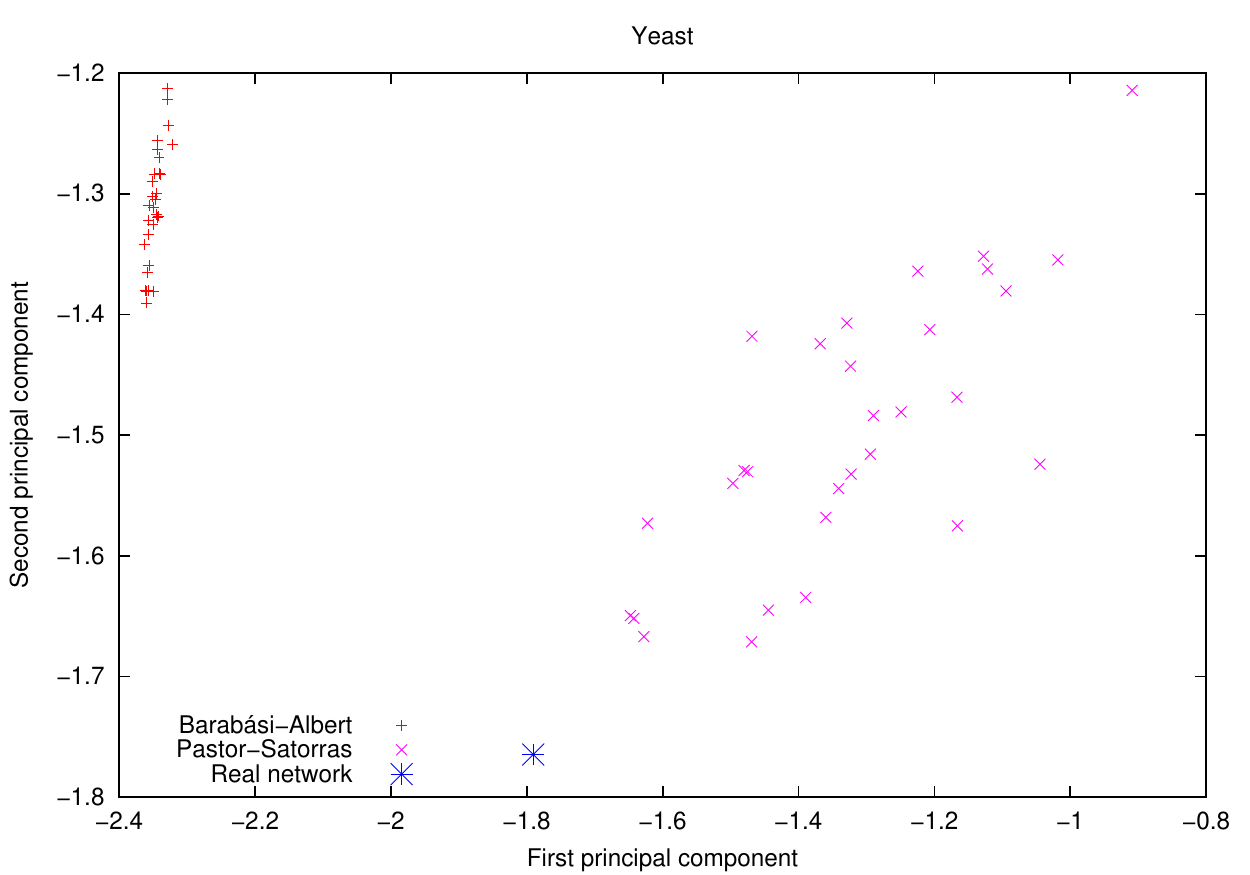}  
  \caption{PCA projections of the positions in the centrality
    correlation profile space of the yeast network and 30 networks
    generated by the Barab\'asi-Albert model and 30 networks generated
    by the Pastor-Satorras et al.\ model.}\label{fig:modeltest}
\end{figure}

Considering the previously presented results, we suggest that the
centrality correlation profile can be used as a tool to test the
adequacy of a network model developed to study a given real network.
If the real network can be considered typical, with respect to the
correlation profile, in comparison to networks generated using the
proposed model, the model is appropriate.  In an ideal case, we would
know the distribution of points representing the generated networks in
the correlation profile space and use standard statistical methods to
evaluate the probability of the real network being generated by the
model.  In practice, when the correlation profile of the model is not
known, we can use PCA projections of the real network and a large
number of generated networks to achieve an informal confirmation of
the model.  To demonstrate this procedure, we use the yeast protein
interaction network from Ref.~\cite{Bu03} and compare it with
Barab\'{a}si-Albert networks and the model developed by
Pastor-Satorras et al.~\cite{Pastor03} specifically for protein
interaction networks.  Figure~\ref{fig:modeltest} shows a PCA
projection of the centrality correlation profile of the network and
30~random networks generated by each model.  The real network is much
closer to the networks generated by the Pastor-Satorras et al.\ model
than to the ones generated the Barab\'{a}si-Albert model.  But the
yeast network cannot be considered a typical network from the
Pastor-Satorras et al.\ model, as it lies outside of the region of
correlation profile space spanned by the random networks generated
according to the model, demonstrating that there are still important
structural details in the real network not accounted for by the model.

\section{Conclusion}
\label{sec:concl}

Various centrality measurements are commonly used to discriminate
important nodes in complex networks.  The different measurements
correspond to different definitions of the importance of the nodes,
but our results have shown that they are in general strongly
correlated for real networks, and even more for the two network models
studied.  We considered the following measurements: degree, closeness,
betweenness, eigenvector, subgraph, current flow closeness, and
current flow betweeness centralities.  For most pairs of centralities,
their Pearson correlation coefficients are above 0.5 for most
networks, with some pair showing coefficients above 0.95 for some
networks, specially the network models.  The log-log scatter plots
show that the correlations are specially strong for high centrality
nodes, where they follow a power law.  But the correlation values vary
strongly from one network to another.  For example, while the Pearson
correlation coefficient between closeness and eigenvector centrality
is 0.99 for the ER model and 0.92 for the karate club network, it is
almost zero for the power grid network.  We proposed therefore the use
of the \emph{centrality correlation profile}, consisting of the values
of the correlation coefficient for all pairs of centralities studied,
to characterize a network.  Our results show that the networks can be
distinguished using this profile.  We have also shown, using the
example of the yeast protein interaction network, how the centrality
correlation profile can be used to verify to what extent a model (in
our example the Pastor-Satorras et al.\ model) is adequate to explain
a given network.

Interesting open questions suggested by this work include: Why are the
correlation coefficients for the network models so strong for almost
all pairs of centralities?  Are the power laws seen for high
centrality values due to specific topological features of the
considered networks or do they result from the definitions of the
measurements?  Why are correlations in real networks consistently
smaller than in the models?  Do the results hold for other models and
real networks?  What kind of topological features makes some
correlations smaller and other larger for a given network?  An answer
to the last question would help us design more adequate models for
some network and therefore understand them better.

\bibliographystyle{unsrt}
\bibliography{correlations}

\end{document}